\documentclass[sigconf]{acmart}

\usepackage{algorithm}
\usepackage{algpseudocode}
\usepackage{subcaption}
\usepackage{graphicx}
\usepackage{caption}

\usepackage{multirow}

\AtBeginDocument{%
  }

\copyrightyear{2025}
\acmYear{2025}
\setcopyright{rightsretained}
\acmConference[SIGSPATIAL '25]{The 33nd ACM International Conference on Advances in Geographic Information Systems}{October 29-November 1, 2025}{Atlanta, GA, USA}
\acmBooktitle{The 32nd ACM International Conference on Advances in Geographic Information Systems (SIGSPATIAL '25), October 29-November 1, 2025, Atlanta, GA, USA}
\acmDOI{10.1145/3678717.3691280}
\acmISBN{979-8-4007-1107-7/24/10}

\begin{document}


%
\title[Modeling Heterogeneity across Varying Spatial Extents]{ Modeling Heterogeneity across Varying Spatial Extents: Discovering Linkages between Sea Ice Retreat and Ice Shelve Melt in the Antarctic }

\author{Maloy Kumar Devnath}
\email{maloyd1@umbc.edu}
\affiliation{%
  \institution{Institute for Harnessing Data and Model Revolution in the Polar Regions, University of Maryland, Baltimore County}
  \streetaddress{1000 Hilltop Circle}
  \state{Maryland}
  \country{USA}
  \postcode{21250}
}

\author{Sudip Chakraborty}
\email{sudipc1@umbc.edu}
\affiliation{%
  \institution{Institute for Harnessing Data and Model Revolution in the Polar Regions, University of Maryland, Baltimore County}
  \streetaddress{1000 Hilltop Circle}
  \state{Maryland}
  \country{USA}
  \postcode{21250}
}

\author{Vandana P. Janeja}
\email{vjaneja@umbc.edu}
\affiliation{%
  \institution{Institute for Harnessing Data and Model Revolution in the Polar Regions, University of Maryland, Baltimore County}
  \streetaddress{1000 Hilltop Circle}
  \state{Maryland}
  \country{USA}
  \postcode{21250}
}

\renewcommand{\shortauthors}{Devnath et al.}

\begin{abstract}

Spatial phenomena often exhibit heterogeneity across spatial extents and even in proximity. This spatial variation is complex to model, especially for large spatial extents that may be varying, for example, ice shelves and sea ice. In this paper, we address this gap and, in particular, highlight its use in understanding linkages between sea ice retreat and Antarctic ice shelf (AIS) melt in the Antarctic. The Antarctic is losing ice at an unprecedented rate. While the role of atmospheric forcing and basal melting on both sea ice retreat and continental ice mass loss has been widely studied, how do retreat of sea ice affects the AIS mass loss has not been investigated. In fact, the linkage between the two has not been well established yet. Traditional models often treat sea ice and AIS as independent systems, limiting their ability to capture localized linkages and cascading feedbacks. To address this, we propose Spatial-Link, a novel graph-based modeling framework that quantifies spatial heterogeneity across varying spatial extents to capture the linkages between sea ice retreat and AIS melt. Our approach constructs a spatial graph using Delaunay triangulation of satellite-derived ice change matrices, where nodes represent regions of significant changes and edges encode proximity and directional consistency. We extract and statistically validate linkage paths using breadth-first search and Monte Carlo simulations to identify significant spatial linkages across varying extents. Experimental results reveal non-local and spatially heterogeneous patterns of coupling, suggesting that sea ice loss can initiate or amplify downstream melting of AIS. Our analyses show how sea ice retreat evolves over an oceanic grid and gradually progresses to the ice shelve - establishing a direct linkage. To the best of our knowledge, this is the first-ever direct linkage methodology between sea ice retreat and AIS melt events. By integrating dynamic cryospheric processes into a unified framework, Spatial-Link offers a scalable, data-driven tool for improving the accuracy of sea-level rise projections and informing targeted climate adaptation strategies.

\end{abstract}




\keywords{Sea ice retreat, Antarctic, Spatial heterogeneity, Delaunay triangulation, Sea-level rise}



 \maketitle

\section{Introduction}

Tobler's first law indicates that everything is related to everything else, with nearby things more related than distant things \cite{walkikar2024discovery}.  However, the associations or linkages between spatial locations can be proximal or far apart, both in time and space, and exhibit spatial heterogenities that affect the nature of such relationships. Modeling this heterogeneity is a key challenge in spatial data science. In addition, modeling similar behaviors or linked behaviors in the presence of heterogeneity is even more challenging~\cite{peters2017elements}. This spatial variation is complex to model, especially for large and varying spatial extents where multiple phenomena may be at play, such as  ice sheets (land) and sea ice.

In this paper, we address this gap and propose a novel method to detect the spatial linkages between different spatial locations across varying spatial extents. Here, linkage refers to interactions or associations between varying spatial extents, which may be influenced by disparate phenomena. We demonstrate our approach in two varying spatial extents,  sea ice retreat and Antarctic ice shelf (AIS) melt, which are governed by two different underlying phenomena that have been intensely researched individually but the linkages have not been well studied. 

Recent studies show that the mass loss of the Antarctic ice sheet is high near the coastal regions~\cite{smith2020pervasive}, AIS is in contact with the warm sea surface temperature and air temperature is higher due to the low elevation and suggests a competing linkage between AIS melt and sea ice retreat~\cite{smith2020pervasive,park2024amundsen_13}. Here we explore the linkages between the sea ice retreat events and the ice shelve that are in contact with each other and proximal to the sea ice. We do not explore the linkages between sea ice retreat events and the AIS that are more inland and far from the ocean. 

Sea ice plays a key role in retreat of AIS. For example one study highlights the role of anomalous atmospheric and sea-ice circulation in the abrupt recessions of the Larsen A and B ice shelve~\cite{christie2022antarctic_14}. Another recent study shows that the reflected solar radiation from Antarctic Sea ice to space increased between 1991 and 2015 but started decreasing after 2016 due to surface albedo reduction in response to the reduced sea ice area~\cite{chakraborty2025impact}. These findings highlight the importance of sea ice loss on AIS. The rationale behind our choice to explore the linkages between AIS melt and sea ice retreat is because of the record sea ice minima (reached a record minimum of 1.965 million $km^{2}$ in 2023 compared to  7.7 million $km^{2}$  in 2014) observed in the Antarctic region~\cite{raphael2022new_2} in recent times and its coincidence with the unprecedented AIS melt~\cite{nasa_ice_sheets_3}. The cryospheric melting in the Antarctic region can have adverse effects on human civilization as 45 percent of the global population lives near the coastline (examples include cities like Los Angeles, Miami, New York), and AIS has the potential to raise the sea level by 200 feet when it melts completely. Currently, they are melting at a rate of 150 gigatons per year~\cite{nasa_ice_sheets_3}. However, the continent has lost an estimated 1200 gigatons of ice between February 2017 and February 2022. In particular, the retreat of grounding lines, especially in the West Antarctic ice shelf, has been interpreted as a critical threshold. There is a knowledge gap regarding the linkages between sea ice retreat and AIS ~\cite{smith2020pervasive,christie2022antarctic_14,eayrs2021rapid,roach2020antarctic}, and significant uncertainty remains in AIS melt estimation and their representation in models and future projections~\cite{eayrs2021rapid,roach2020antarctic}. In addition, the impact of the blanketing loss of sea ice is not well understood.

To address this gap, we propose {\it Spatial-Link}, a graph-based modeling framework designed to quantify heterogeneous spatial linkages between varying spatial phenomena. In particular, we demonstrate linkages between sea ice retreat and AIS melt acceleration. This approach conceptualizes distinct ice regions as interconnected nodes. Our approach allows for a more nuanced representation of spatially varying processes and feedback even in proximity across large and varying spatial extents where multiple complex phenomena are at play. Traditionally, association rule mining is very popular in detecting the linkages. For example, associations have helped researchers understand the causes behind subjects’ health-related issues and for better guidance of the subjects for solving their health issues~\cite{altaf2017applications}. Previously, Tolu et al. have used the association rule mining method to detect the association between ice sheet depth and runoff with the meteorological factors over the Greenland area in the Arctic~\cite{ale2023multi}. However, they have been conducted and detected linkages between meteorology, such as temperature, and ice sheet melting over the same grid. They explored how changes in the meteorological parameters affect the ice sheet melting. However, in this paper, we are considering the linkages between two grids that are far apart. Sea ice grids are located over the Southern Ocean, whereas the AIS grids are located over the continent. We are evaluating how a grid experiencing sea ice retreat far from the coast affects a sea ice grid (over the ocean) next to it, and by doing that, the retreat gradually propagates towards the ice shelve and affects ice shelf melting.

By resolving these localized linkages, Spatial-Link can help localized identification of the sea ice retreat that may contribute to potentially increased AIS melt. Our approach can detect the cascading effects that are not captured in conventional models, such as how sea ice melt propagates the ice shelf's melt.  As efforts to constrain future sea-level rise intensify, frameworks like Spatial-Link can provide valuable insights for improving future sea level rise projections and informing targeted climate adaptation strategies. Specifically, we make the following contributions:

\begin{itemize}
    \item We propose Spatial-Link, a graph-based novel framework that models spatially heterogeneous linkages between sea ice retreat and AIS melting.
    \item Our framework uncovers directional dependencies, demonstrating how sea ice retreat drives downstream AIS melt.

    \item We further validate Spatial-Link$_{Benchmark}$ using MERRA-2 (a global atmospheric reanalysis dataset from NASA) sulfate aerosol data at the Gates of the Mountain station in Montana, showing eastward pollution transport from wildfire-induced PM$_{2.5}$ spikes.

\end{itemize}

The rest of the paper is organized as follows. Section \ref{related} describes related work. Section~\ref{sec:method} outlines our methodology. In Section~\ref{sec:experiment}, we present experimental results. Finally, we present conclusions and future work in Section~\ref{sec:con}.

\section{Related Work} \label{related}
Graph theory works as a powerful analytical tool for detecting linkages among multiple entities by modeling them as nodes and their linkages as edges~\cite{friedman1983graph}. In parallel, association rules offer interpretable rule-based insights and have been widely adopted in many domains. 

Together, graph-based models and association rules provide both structural and semantic perspectives for understanding linkages in complex and high-dimensional datasets. In environmental and climate domains, graph theory has been employed to examine the controllability and heterogeneity of river networks~\cite{sarker2025controllability}, and to analyze the varying impacts of agriculture on the water–energy balance across different climatic regions~\cite{zaerpour2025agriculture}. Similarly, association rules have proven valuable in detecting complex relationships between large-scale atmospheric teleconnections and extreme precipitation events~\cite{abdelmoneim5084852data}, improving weather forecasting models~\cite{kumar2025analysis}, and advancing spatiotemporal forecasting approaches for climate prediction~\cite{deng2025hybrid}.

Existing studies on sea ice retreat and AIS melt have largely approached these processes in isolation. They have focused on their respective thermodynamic and dynamic behaviors. For instance,~\cite{Feltham2006} conceptualizes sea ice as a mushy layer, a perspective echoed by ~\cite{Singh2021}, who has emphasized the importance of modeling sea ice as a complex mixture of solid ice and brine inclusions. Advanced numerical methods have also been used to model sea ice behavior. For example,~\cite{Schwarz2021} utilizes finite element methods and data assimilation to simulate sea ice drift under realistic wind conditions.  However, these studies treat sea ice as an isolated system. In contrast, ice sheet modeling has primarily focused on long-term mass balance and dynamic responses to climate drivers. Sea ice responds rapidly to seasonal changes, while AIS evolves more slowly and interacts with climate through longer-term feedback mechanics in AIS melts~\cite{deconto2007sea}. As a result, research has tended to treat these systems separately, despite growing evidence that their linkages can significantly amplify polar climate responses.

Fortunately, some research has begun to examine linkages between sea ice and AIS. Rennermalm et al. highlight a strong covariance between sea ice retreat and surface melting on the Greenland ice sheet, particularly during late summer~\cite{rennermalm2009does}. They suggest that reductions in offshore sea ice warm the ocean’s mixed layer, facilitating the onshore advection of heat and raising air temperatures over the AIS -factors that significantly increase surface melt probabilities. However, this study is limited to the Arctic and does not extend to a global understanding of cryosphere coupling.~\cite{pritchard2012antarctic} identifies changes in wind forcing as a driver of warm ocean water intrusion beneath West Antarctic ice shelve. Such interactions or linkages may not occur within the same spatial footprint, they can vary in strength and directions, both spatially and with time, making them highly heterogeneous. Wind forcing is leading to dynamic thinning of grounded glaciers. This coupling between ocean-driven ice shelf thinning and glacier acceleration highlights the complex mechanisms behind Antarctic mass loss. Likewise,~\cite{smith2020pervasive} emphasizes the role of ocean and atmospheric processes in accelerating mass loss from both Greenland and AIS. Thermodynamic influences such as basal melting and subglacial discharge further complicate ice sheet responses. Similarly,~\cite{himmich2024} discusses how sea ice thinning and earlier retreat increase residual heat release, thereby impacting adjacent ice masses. These studies suggest a possibility of numerous linkages over the Southern Ocean and the Antarctic Continent. However, no study has ever tried to focus on detecting the linkages between them, primarily due to the fact that there might be numerous linkages that warrant analyses using a large volume of satellite datasets and images. Traditional computational methods, such as statistical analyses, may struggle to detect linkages effectively and are often time-consuming and computationally expensive. In contrast, machine learning techniques can be suitable for this type of problem. 

As mentioned before, the association rules are a popular data mining technique used to uncover interesting relationships, patterns, or linkages among variables in large datasets. It has been originally introduced for market basket analysis, where it identifies sets of products that frequently co-occur in customer transactions. In the current context of the paper, it has been used in visualizing spatiotemporal climate data~\cite{wang2018association}; however, it often struggles to detect and evaluate the linkages due to the heterogeneous nature of the problem, that are associated with problems related to the geospatial data that are always heterogeneous and big in volume. 

Some researches show the utility of the graph theory that provides a flexible and powerful representation for association rules, where items are modeled to depict connected nodes and the relationships therein (e.g., frequent co-occurrences or rules)  are represented as edges~\cite{shanavas2020graph}. This graph-based approach allows mining algorithms to uncover not only direct linkages but also more complex and higher-order patterns through graph traversal, subgraph mining, and community detection~\cite{janeja2008random}. It is especially effective in large, complex datasets such as social networks, bioinformatics, and e-commerce~\cite{li2020hierarchical}, where traditional itemset-based rule mining might miss indirect or weakly connected patterns. However, the use of graph theory to show linkages is still underexplored, especially in the field of cryosphere science, which is tremendously important to understand in more detail than in its current form, as we mentioned before. Sea ice cover protects the ice shelve (in the coastal area), which is further a blanket to the Antarctic ice sheets in the interior of the continent, from thermal forces like temperature and mechanical forces like wind and wave actions. We hypothesize that the graph theory will be suitable for identifying linkages between two varying spatial extents that can be spatially far apart and heterogeneous, both in terms of their spatial distribution and physical nature of retreating or melting.

\section{Methodology}
\label{sec:method}
\begin{figure*}[thb!]
    \centering
    \includegraphics[width=2\columnwidth]
    {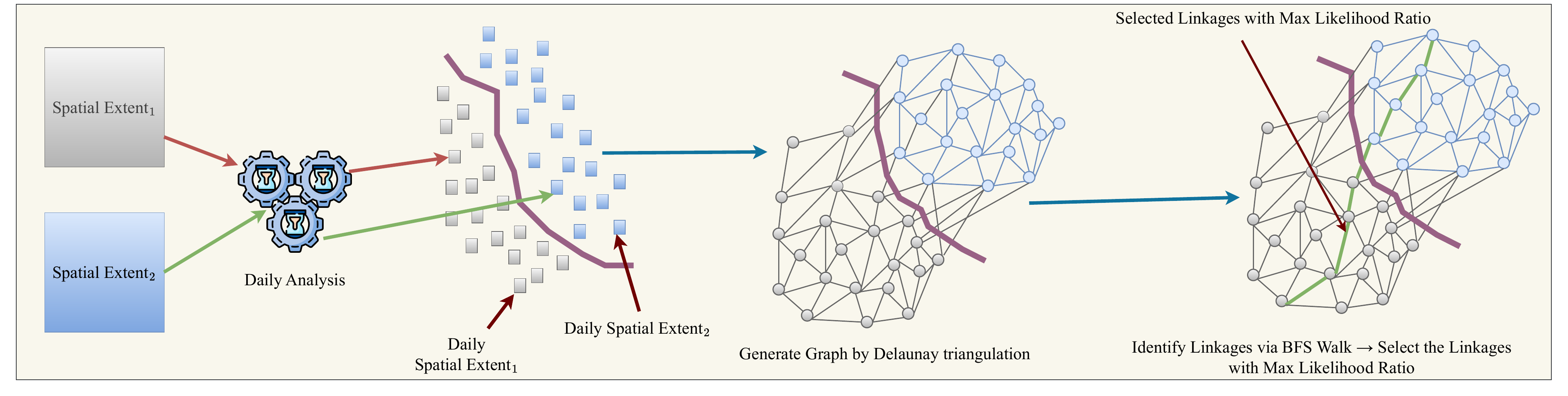}

   \caption{Overview of the Spatial-Link framework for detecting statistically significant linkages between spatial extent$_{1}$ and spatial extent$_{2}$. The process includes Delaunay triangulation for spatial adjacency, graph construction with weighted or non-weighted edges, BFS-based linkage extraction, and Monte Carlo testing for detecting significant linkages.}

    \label{fig:overall_method}

\end{figure*}

Figure~\ref{fig:overall_method} illustrates the overall framework used to identify linkages between sea ice retreat and AIS melt in the Antarctic region. The process begins by detecting regions of significant change using satellite-derived matrices of sea ice retreat events, along with corresponding regions of sea ice changes in the neighborhood. Then, determining whether the change is continuous enough to reach the ice shelve. These change points are then spatially connected using Delaunay triangulation to build a graph that captures proximity-based relationships~\cite{janeja2008random}. Edges are formed only between geographically close regions (within \textasciitilde
275 km) that satisfy specific thresholds for physical change. This corresponds to continuous 11 nodes along in the latitudinal direction to find the graphs. The selection of the \textasciitilde
275 km threshold is motivated by the fact that grid cells need to be proximal enough to allow for potential linkages or linkages~\cite{smith2020pervasive}. Each graph may vary, can exhibit even a shorter length, and is discussed in the results. This distance is gauged from previous studies that observed the alignment of sea ice retreat and AIS melting in proximity~\cite{smith2020pervasive}. These studies did not explore the linkages, but showed the side-by-side positioning of sea ice retreat and AIS melt. Each edge is assigned a weight based on whether the directions of change (sea ice retreat or melt for AIS) are consistent between the connected points~\cite{janeja2008random}. Using breadth-first search (BFS)~\cite{li2025extracting}, we extract all valid short paths that connect sea ice retreat points to AIS melting regions. We also examine whether the paths are statistically significant. To do so, we perform Monte Carlo simulations to generate null distributions and compute corresponding p-values~\cite{janeja2008random}. Paths with p-values below 0.05 are considered statistically significant, indicating potential physical linkages between sea ice retreat and AIS melt.

\begin{algorithm}
    \caption{Spatial-Link}
    \label{algo:significant_paths}

    \begin{algorithmic}[1]
    
    \Statex \textbf{Input:} Sea ice change matrix $SIC$, Ice shelve change matrix $AIS$, threshold ranges $T_{sic}$ and $T_{ais}$, maximum path length $L$, number of Monte Carlo samples $M$
    \Statex \textbf{Output:} Statistically significant linkage paths between sea ice retreat and AIS melting

    \State Identify sea ice retreat locations from $SIC$ and AIS melting locations from $AIS$
    \State Build Delaunay triangulation over all valid points
    \State Initialize an empty graph $G$

    \For{each edge $(u, v)$ in triangulation}
        \If{distance between $u$ and $v \leq 11$}
            \State Extract $SIC$/$AIS$ change values for $u$ and $v$
            \If{$u$ and $v$ pass threshold filters $T_{sic}$ and $T_{ais}$}
                \State Set edge weight: $+1$ if signs of changes match, $-1$ otherwise
                \State Add $(u, v)$ to graph $G$
            \EndIf
        \EndIf
    \EndFor

    \State Initialize path set $P$
    \For{each node $s$ in sea ice retreat points}
        \State Use BFS to find all valid paths of length $\leq L$ ending in AIS melting locations
        \State Add all valid paths to $P$
    \EndFor

    \State Initialize null distributions using $M$ Monte Carlo simulations
    \For{each path $p$ in $P$}
        \State Compute observed positive correlation score of $p$
        \State Compare with null distribution to calculate $p$ value
        \If{$p$value $< 0.05$}
            \State Mark $p$ as significant
        \EndIf
    \EndFor

    \State Return all significant paths

    \end{algorithmic}
\end{algorithm}

\subsection{Spatial-Link}

The Spatial-Link (Algorithm~\ref{algo:significant_paths}) starts by identifying key sea ice retreat and AIS melt locations from the sea ice concentration change matrix ($SIC$) and the AIS change matrix ($AIS$), which is shown in line 1. A Delaunay triangulation is then constructed over all the valid points to efficiently establish spatial adjacency relationships. Line 2 depicts the Delaunay triangulation operation in our proposed method. An empty graph $G$ is initialized to store spatial linkages or paths, which is shown in line 3.

For each edge $(u, v)$ in the triangulated mesh, the proposed approach checks whether the Euclidean distance between the two points is less than or equal to 11 grid cells, which corresponds to 275 km, considering each grid cell represents 25 km. These criteria are shown in lines 4 to 5. If this distance criterion is met, the $SIC$ and $AIS$ change values for both $u$ and $v$ are extracted. These values are passed through threshold filters $T_{sic}$ and $T_{ais}$ to ensure that only meaningful changes are considered. If both nodes meet the threshold criteria, the method assigns a weight of $+1$ to the edge if the signs of change at $u$ and $v$ match, and $-1$ if they differ. The validated edge is then added to the linkage graph $G$. These operations are shown in lines 6 to 12.

Next, a path set $P$ is initialized to store candidate linkage paths, which is shown in line 13. For each node $s$ identified as a sea ice retreat point, a BFS is conducted to find all valid paths of length up to $L$ that terminate in ice shelve' melt locations. All such paths are then added to $P$. These operations are shown between lines 14 and 17.

To determine statistical significance, the method constructs a null distribution using $M$ Monte Carlo simulations, which is shown in line 18. For each path $p \in P$, it computes an observed positive correlation score and compares this score against the null distribution to obtain a $p$ value. If the $p$ value is less than $0.05$, the path is considered statistically significant. These operations are shown between line 19 and 25.

Finally, the proposed method returns all statistically significant paths, which represent meaningful spatial linkages between sea ice retreat and AIS melting regions which as shown in line 26.

\subsection{Mapping Sea Ice Retreat and AIS Melt Linkages with CMAD based Spatial-Link}

The Spatial-Link has been modified for our Convolution Matrix
Anomaly Detection (CMAD)~\cite{devnath2024cmad}-based Spatial-Link, which analyzes anomalous sea ice retreat and AIS melting. In this adaptation, the method processes the sea ice retreat matrix and the AIS melt matrix to identify statistically meaningful spatial linkages, in context with the anomalous sea ice retreat events identified and predicted by CMAD~\cite{devnath2024cmad,devnath2024deep}. Specifically, in lines 7–8 of Spatial-Link~\ref{algo:significant_paths}, if both nodes $u$ and $v$ correspond to anomalous sea ice retreat regions and satisfy the threshold condition $T_{ais}$ for AIS mass change, an edge is assigned a weight of $+1$ to indicate a reinforcing linkage. Otherwise, a weight of $-1$ is assigned to capture a diverging AIS or sea ice melting change pattern.

\subsection{Spatial-Link$_{Benchmark}$ for Benchmark AAR Data}

For verifying our Spatial-Link, we have modified our graph-based approach to model the spatial dynamics of sulfate aerosol atmospheric rivers (AAR) that carry pollutants from their source region (California wildfires) to a receptor and pristine region (Montana, United States). A study shows an elevation of particulate matter with a diameter less than 2.5 microns, which tremendously affects our respiratory system, including lungs ~\cite{xing2016impact}, in a station in Montana that is far from the source region.  This approach is integrated into our Spatial-Link$_{Benchmark}$.

\subsubsection{Graph Construction Using Delaunay Triangulation}

We have constructed spatial graphs where each node represents elevated sulfate aerosol observation point derived from MERRA-2 Sulfate AAR data~\cite{chakraborty2022aerosol}. First, we follow the exact physical description mentioned in the algorithm available here~\cite{AAR_22}. Then we apply the Delaunay triangulation method that preserves local spatial relationships while avoiding artificial connections across distant points~\cite{janeja2008random}. Once the full graph is constructed, we extract connected components and retain only those with a length of more than 2000 km to eliminate small or noisy clusters, as described in the definition of atmospheric rivers~\cite{guan2015detection}. For each component, we compute a geodesic-inspired spatial extent using a latitude longitude-corrected distance metric to follow the Euclidean distance that varies longitudinally, but not latitudinally, with a close approximation of 111.11 km per degree with longitude correction (cosine of latitude), as the earth is a sphere. Components larger than 2000 km are included in the final graph, as in the original AAR methodology and the algorithms.


To ensure the constructed graphs reflect physically meaningful transport structures, we validate them using data from the IMPROVE network~\cite{green2009comparison} particulate matter 2.5, which are tiny aerosol particles smaller than 2.5 micron in size and are extremely harmful for lungs~\cite{xing2016impact}, measurements~\cite{green2009comparison}, confirming alignment with observed aerosol transport patterns~\cite{chakraborty2022aerosol}. This validation confirms that the graph topology captures realistic atmospheric pathways. Specifically, for December 2014 data, our approach preserves over 96\% of the original spatial structure, on average of 30 days, in detecting the AARs over a domain of North America, including the contiguous United States, southern Canada, and parts of the North Atlantic. The high accuracy in detecting the AAR by our methodology shows the feasibility of using graph theory for detecting spatiotemporal linkages. These results demonstrate the reliability and fidelity of our graph-based modeling method for sulfate AAR dynamics and show it applicability in the data that are spatially heterogeneous.

\section{Experiments}
\label{sec:experiment}

Table~\ref{tab:datasets} shows the datasets and images used in this study. Before applying our Spatial-Link, we have standardized the resolution of the input datasets to ensure compatibility. The Sea Ice Concentration (SIC) data, initially at a $332 \times 316$ grid, has been resampled to a $720 \times 1440$ grid corresponding to a $0.25^\circ \times 0.25^\circ$ spatial resolution. This has allowed integration with other geospatial datasets and enhanced spatial detail. The SIC data is highly heterogeneous and grids can have SIC concentration ranging between 1 (grid fully covered by ice) to 0 (no ice in the grid, only oceanic water). For CMAD-based anomaly detection, each $332 \times 316$ grid cell has been labeled as anomalous ($1$) or non-anomalous ($0$) before projection onto the higher resolution grid. The AIS melt dataset has already been available at $720 \times 1440$ resolution. We do not require resampling. We have restricted our analysis to the Antarctic region, selecting latitudes from $90^\circ$ to $60^\circ$S and longitudes from $130^\circ$ to $30^\circ$W, corresponding to grid indices $0$ to $120$ (latitude) and $200$ to $600$ (longitude) in the $0.25^\circ$ grid system. We have focused our analysis on the region spanning \(90^\circ\) to \(60^\circ\)S latitude and \(130^\circ\) to \(30^\circ\)W longitude, as existing literature highlights this portion of West Antarctica as a critical hotspot for accelerated sea ice retreat and AIS melting. We next present the results from our Spatial-Link, CMAD-based Spatial-Link and Spatial-Link$_{Benchmark}$ analyses and discuss their significance and implications.


\begin{table*}[!htb]
\centering
\footnotesize
\caption{Summary of datasets used in this study.}
\label{tab:datasets}
\begin{tabular}{@{}p{3.2cm}|p{3cm}|p{4cm}|p{5.5cm}@{}}
\hline
\textbf{Case Study} & \textbf{Dataset} & \textbf{Satellite Data Sources} & \textbf{Resolution / Temporal / Size} \\
\hline

\multirow{3}{=}{\raggedright Antarctic Sea Ice Retreat and AIS (varying spatial extents)} 
& Sea Ice Extent Images (satellite images) & SMMR, SSM/I, SSMIS, NIMBUS Passive Microwave .png images & 332 $\times$ 316 pixels (\textasciitilde 648 MB) \\
\cline{2-4}
& Sea Ice Concentration & SMMR, SSM/I, SSMIS, Nimbus-7 .nc numerical data & 720 $\times$ 1440, 1 day, (\textasciitilde 36 GB)\\
\cline{2-4}
& Snow Depth Data (AIS)~\cite{munozsabater2019era5land} & ERA5-Land (.nc) & 720 $\times$ 1440, 1 day \\
\hline

\multirow{2}{=}{\raggedright Atmospheric Sulfate Transport for Benchmarking (varying spatial extents)} 
& Aerosol Robotic Network (AERONET) and corresponding Aerosol Optical Depth (AOD) measurements~\cite{green2009comparison} & Dataverse repository~\cite{AAR_22} (.nc) & 576 $\times$ 361 grid, 6-hour timestep, 24 years (1997–2020), (\textasciitilde 677 GB) \\
\cline{2-4}
& Integrated Aerosol Transport (MERRA-2) & MERRA-2 aerosol and meteorological variables~\cite{chakraborty2023aerosol} (.nc) & 576 $\times$ 361 grid, 6-hour timestep, 24 years (1997–2020), ( \textasciitilde 697 GB) \\
\hline

\end{tabular}
\end{table*}












\begin{figure*}[thb!]
    \centering
    \includegraphics[width=2\columnwidth]
    {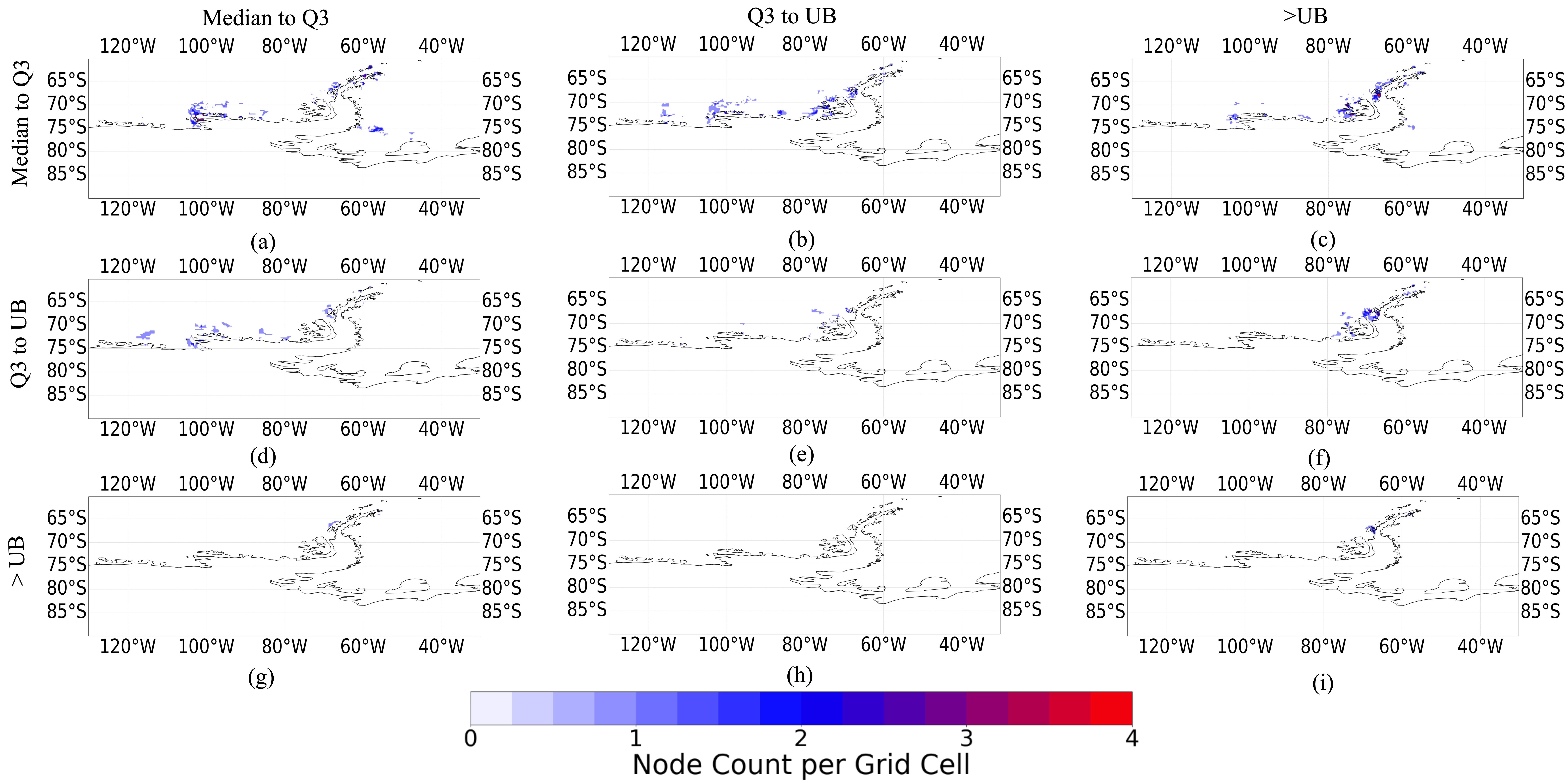}

    \caption{Spatial linkages detected using the Spatial-Link across December 2019. Each subfigure (a)–(i) represents a distinct pairing of sea ice retreat and ice shelve melting threshold levels, showing statistically significant connected components over West and South Antarctica. The figure illustrates how different combinations of moderate, high, and sea ice retreat lead to varying spatial linkages, highlighting dynamic AIS melting.}

    \label{fig:months}

\end{figure*}

\begin{figure*}[thb!]
    \centering
    \includegraphics[width=2\columnwidth]
    {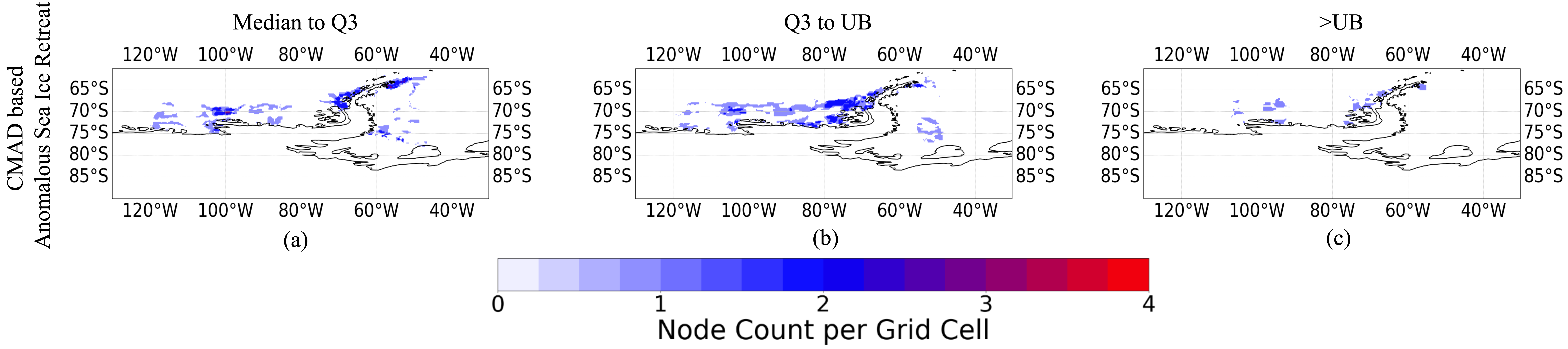} 
   \caption{Spatial linkage detected using the CMAD-based Spatial-Link for December 2019. Each subfigure (a)–(c) corresponds to a different ice shelf melting threshold range (Median–Q3, Q3–UB, >UB) with anomalous sea ice retreat. The figure shows how different levels of AIS melting—moderate, high, and extreme—are influenced by the retreat of sea ice and varying spatial linkages, emphasizing the dynamic nature of this melting process.}

    \label{fig:months_cmad}

\end{figure*}

\subsection{Spatial-Link Results}
To comprehensively investigate the spatial linkages between sea ice retreat and AIS melting, we analyzed nine threshold-based combinations using our Spatial-Link. Specifically, we have categorized both sea ice retreat and AIS melt into three threshold intervals—Median to Q3 (moderate), Q3 to Upper Bound (UB) or high, and greater than UB (anomalous)—resulting in a $3\times3$ grid of possible pairwise linkages which is shown in Figure~\ref{fig:months}. This approach enables a nuanced assessment of how varying intensities of sea ice retreat relate spatially to corresponding AIS melt patterns across a range of severity levels. The results from these combinations offer critical insights into the heterogeneity of coupled cryospheric changes over the West Antarctic region.

\begin{figure}[!htb]
    \centering
    
    \includegraphics[width=\columnwidth]{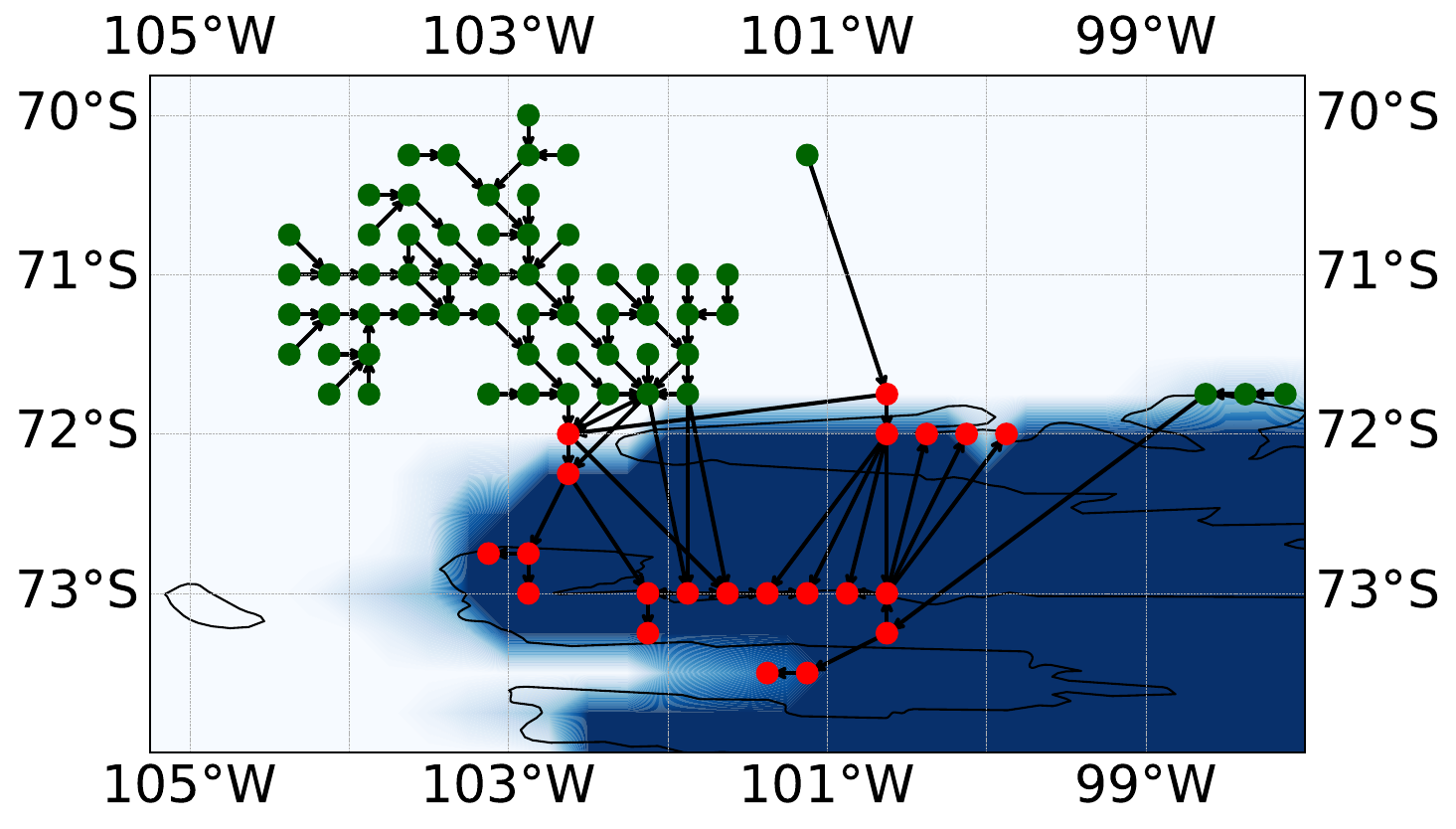}

    \caption{Spatial linkage between sea ice retreat (Median–Q3) and AIS melt (Median–Q3) in the Amundsen Sea sector. Green nodes represent sea ice retreat regions, while red nodes indicate AIS melt regions. 
}
    \label{fig:r1-r1}
\end{figure}


    


\begin{figure}[!htb]
    \centering
    
    \includegraphics[width=\columnwidth]{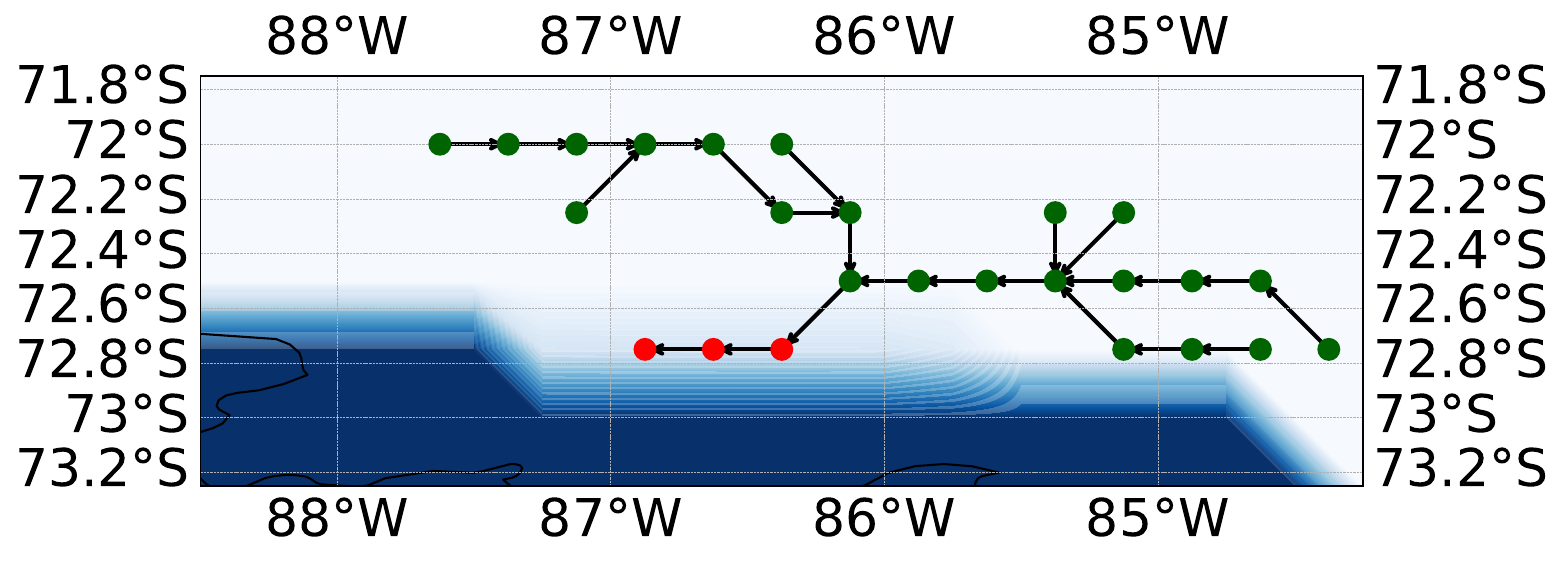}

    \caption{Spatial linkage between sea ice retreat (Median–Q3) and AIS melt (>UB) in the south of the Pine Island Glacier.} 
    \label{fig:r1-r3}
\end{figure}

\begin{figure}[!htb]
    \centering
    
    \includegraphics[width=\columnwidth]{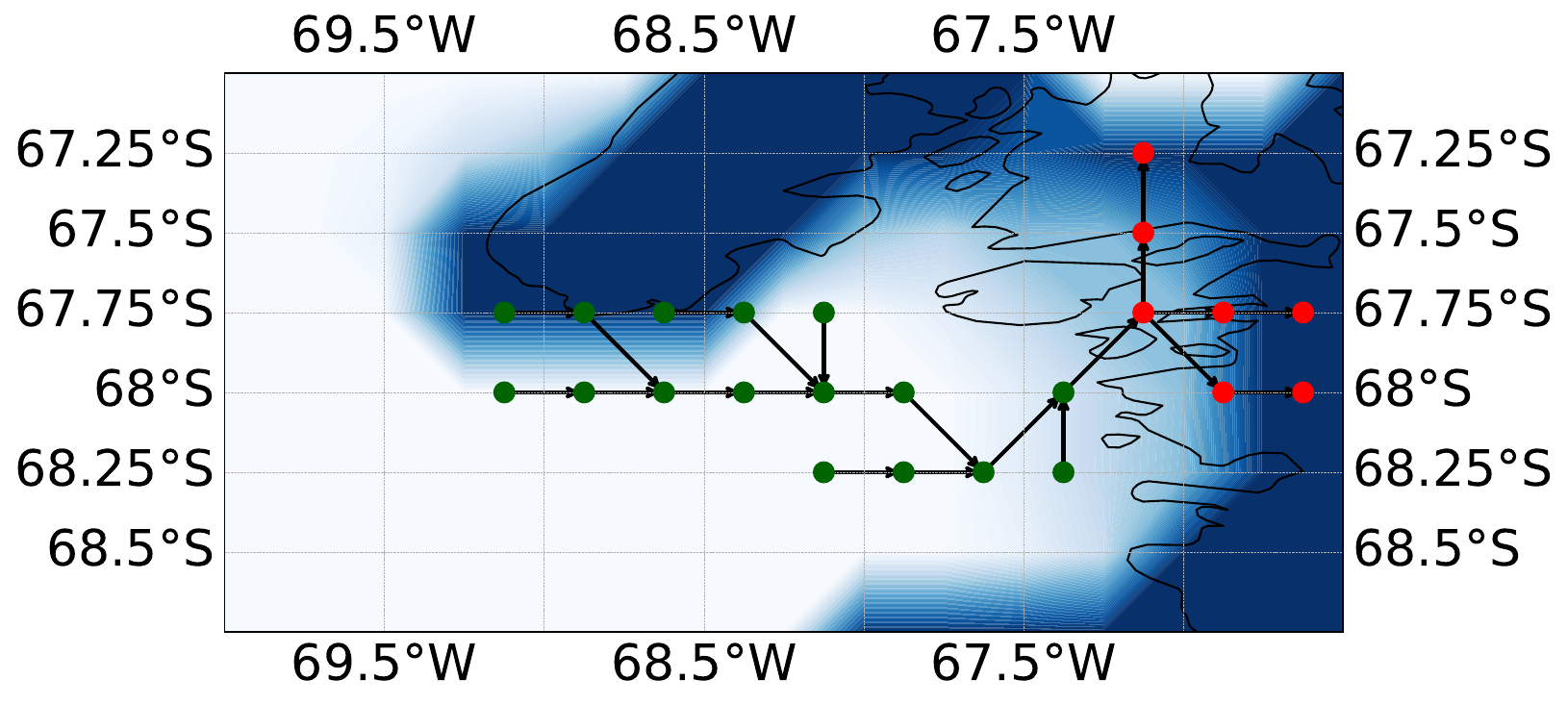}

    \caption{Spatial linkage between sea ice retreat (Q3–UB) and AIS melt (Median-Q3) in the Antarctic Peninsula region.
}
    \label{fig:r2-r1}
\end{figure}

    


    


    


\begin{figure}[!htb]
    \centering
    
    \includegraphics[width=\columnwidth]{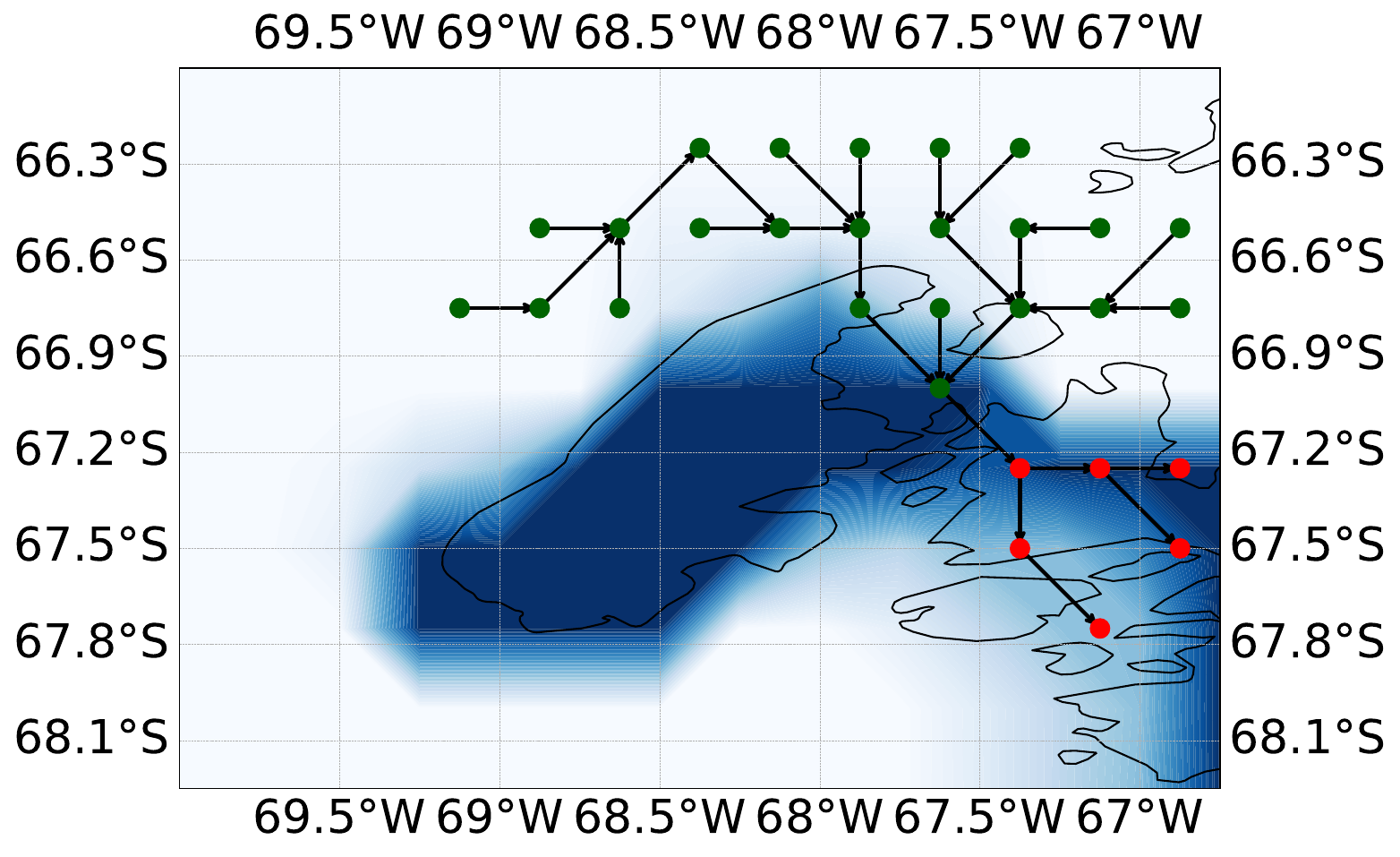}

    \caption{Spatial linkage between sea ice retreat (>UB) and AIS melt (>UB) in the Southwest Antarctic Peninsula, where Palmer Land meets Graham Land.
}
    \label{fig:r3-r3}
\end{figure}

    


\begin{figure}[!htb]
    \centering
    
    \includegraphics[width=\columnwidth]{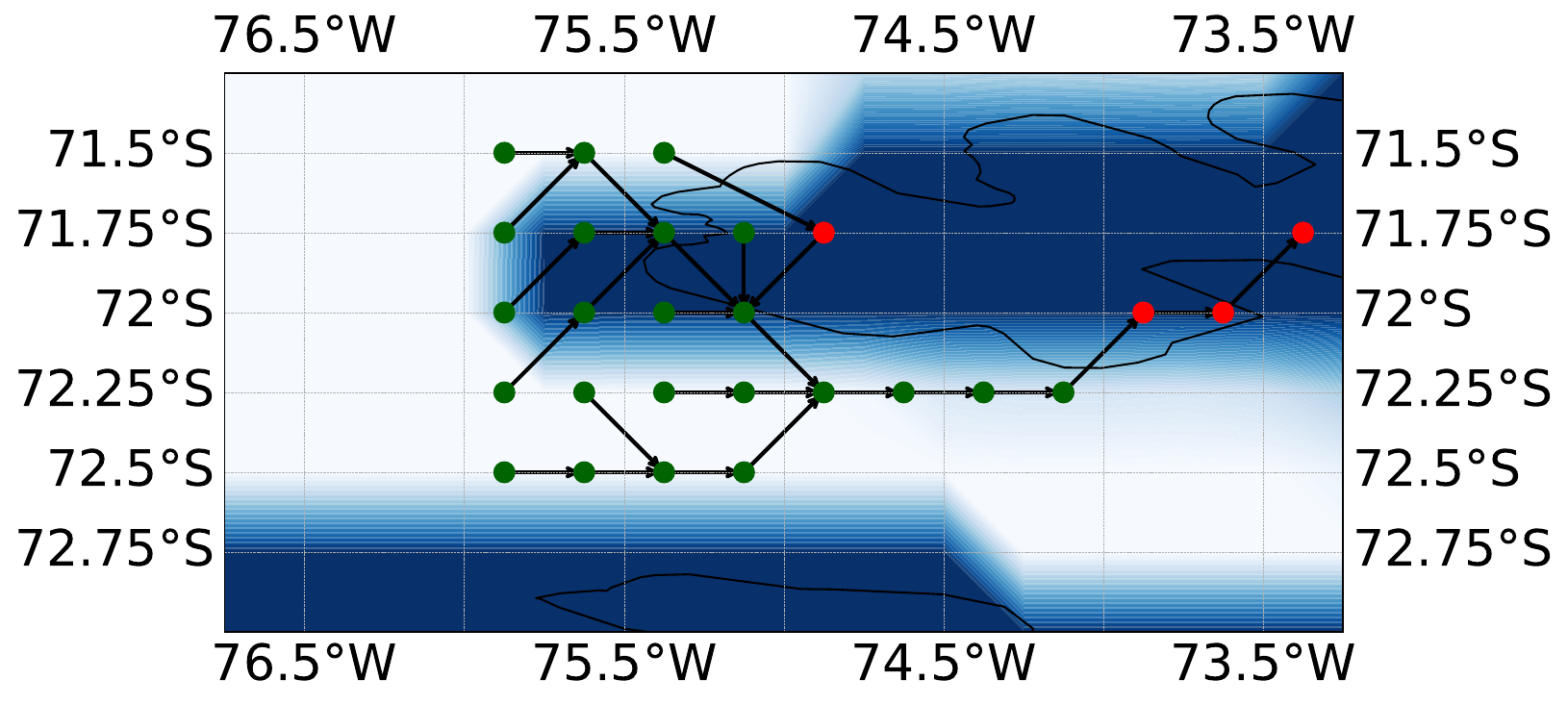}

    \caption{Spatial linkage between anomalous sea ice retreat and AIS melt (Q3 to UB) in West Antarctica, specifically in the Bellingshausen Sea sector near the base of the Antarctic Peninsula.}
    \label{fig:r2}
\end{figure}

\begin{figure}[!htb]
    \centering
    
    \includegraphics[width=\columnwidth]{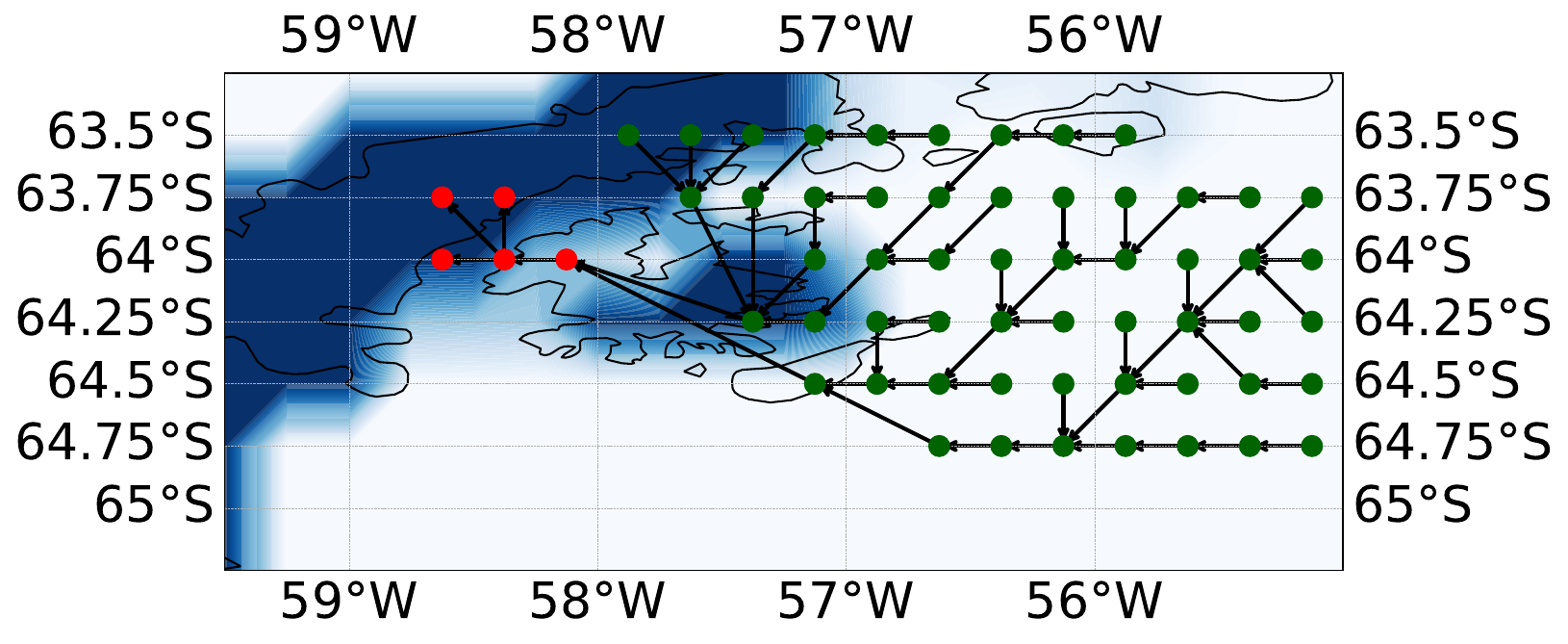}

    \caption{Spatial linkage between anomalous sea ice retreat and AIS melt (>UB) in the northern Antarctic Peninsula area. 
}
    \label{fig:r3}
\end{figure}

\begin{figure}[!htb]
    \centering
    
    \includegraphics[width=\columnwidth]{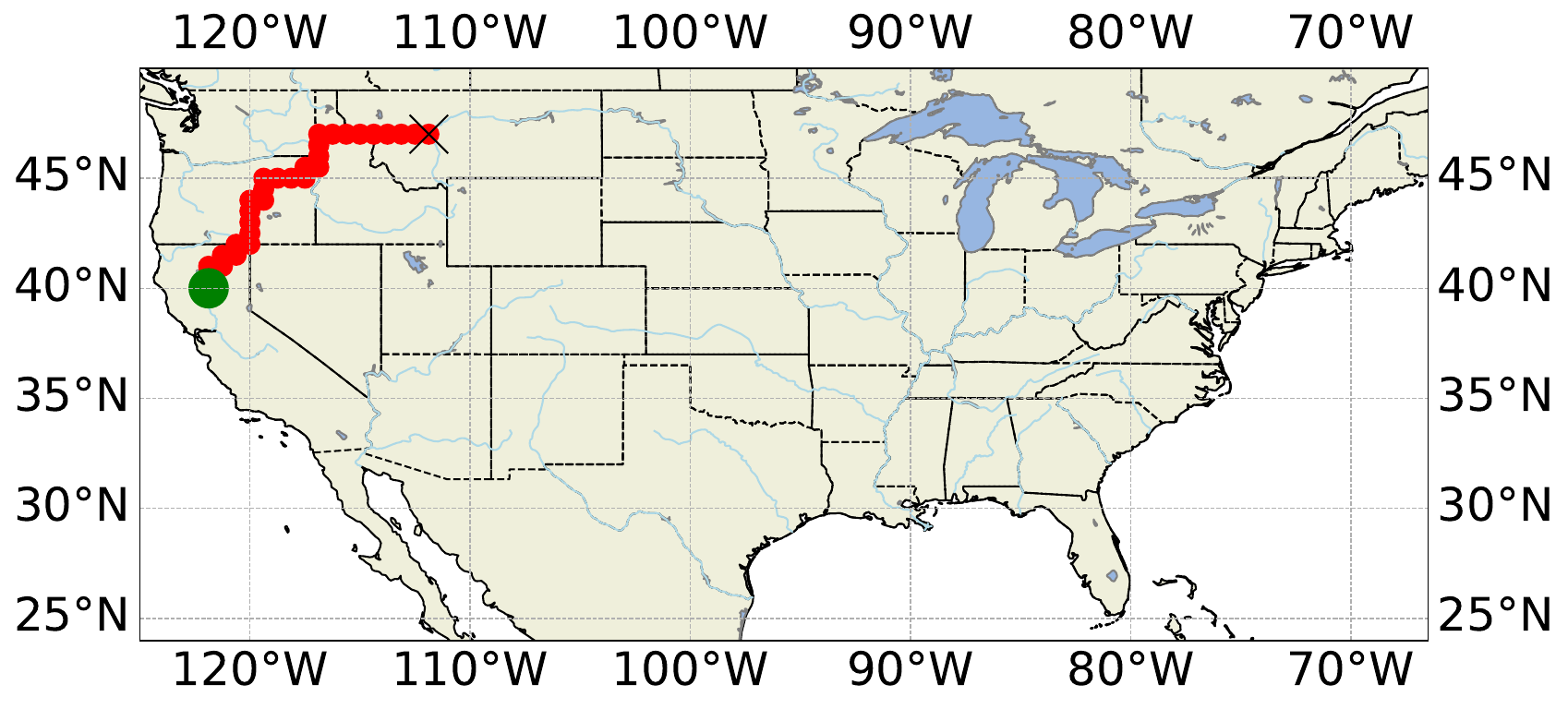}

    \caption{Significant wildfire-driven aerosol transport paths on 20200915 identified by Spatial-Link$_{Benchmark}$, aligning with elevated PM$_{2.5}$ at GAMO1. The green node represents the wildfire origin, the red nodes indicate wildfire propagation, and the black 'X' marks the PM$_{2.5}$ observation location.}
    \label{fig:benchmark}
\end{figure}

\subsubsection*{Sea ice retreat and AIS melt linkage Patterns}

To examine the moderate linkage between sea ice retreat and AIS melting, we have analyzed the spatial link results using the Median-Q3 thresholds for both sea ice retreat and AIS melt across December 2019, which are shown in Figure~\ref{fig:months}(a). Our findings revealed consistent linkage regions throughout the month, particularly concentrated in West Antarctica. These regions span latitudes from approximately 77.25\textdegree S to 61.5\textdegree S and longitudes from 116.25\textdegree W to 46.75\textdegree W, encompassing glaciologically active areas such as the Amundsen and Bellingshausen Seas, Thwaites Glacier, and adjacent inland sectors. Previous studies validate these findings as they also show that the regions stated above are more prone to sea ice retreat and AIS melting~\cite{smith2020pervasive}. This area includes parts of the Thwaites and Pine Island Glacier systems~\cite{thurnherr2014export}, known for their vulnerability to ocean-driven melting and significant contributions to dynamic ice mass loss. A maximum of 4 linakages (red color) in that month has been observed over a grid. This indicates that the sea ice retreat to ice shelve has been linked 4 times through that grid.
To investigate the spatial linkages between moderate sea ice retreat and anomalous levels of AIS melting, we have analyzed spatially linked regions during December 2019 using the Spatial-Link. For this combination (sea ice retreat: Median-Q3, AIS melting: >UB), significant patterns of connected components have been consistently identified, which is shown in Figure~\ref{fig:months}(b). We have observed similar results in how Median-Q3 sea ice retreat affects the Q3-UB (not shown to avoid redundancy) ice shelf melting. It is important to note that the frequency of occurrences of the nodes decreases over the Bellingshausen Sea when we explore the impacts of the Median-Q3 sea ice retreat and anomalous (>UB) ice shelf melting, which is shown in  Figure~\ref{fig:months}(c). However, a localized yet significant linkage cluster has been shown in Figure~\ref{fig:months}(c) between longitudes $71.5^\circ$W and $68.0^\circ$W and latitudes $69.0^\circ$S to $67.0^\circ$S. This region lies along the central Bellingshausen Sea coast, an area subject to intense seasonal variability and ocean-ice-atmosphere coupling~\cite{smith2020pervasive}. This is as expected since in nature, we expect the outliers to occur, but less frequently. However, the recurrence and expansion of these links underscore a strong spatial correlation where even moderate sea ice retreat can coincide with peak-level ice sheet mass loss, suggesting the influence of nonlinear feedback in cryospheric processes. Notably, a spatially linked interaction zone has been observed in  Figure~\ref{fig:r1-r3} between $89^\circ$W and $86^\circ$W longitude and $73.2^\circ$S to $71.8^\circ$S latitude, indicating that moderate sea ice retreat may be closely associated with extreme AIS melting near the inland sector of Pine Island Glacier, further supporting the presence of sea ice retreat impacts on AIS melting. 
For any other combinations, we observed that less frequent linkages occurred in that month for the same reason. These patterns in the Figures~\ref{fig:months}(d-f) suggest that intense, high sea ice retreat links more with anomalous inland melt signatures rather than moderate and high ice shelve melting. Almost no linkages are observed when sea ice retreat is anomalous, which is shown in Figures~\ref{fig:months}(g-i). No statistically significant spatial linkages have been identified between extreme (sea ice retreat: $>$UB) and severe (AIS melting: Q3-UB) during the analyzed periods, which is shown in Figure~\ref{fig:months}(h). This suggests a lack of consistent linkage between the most extreme sea ice retreat and the upper-tier melting responses of the AIS within the study region and timeframe. 
However, these results highlight that moderate (sea ice retreat: Median-Q3) plays a key role in triggering spatial connections with more intense AIS melt responses (AIS melting: Q3-UB, Q3- UB and >UB) which is shown in the Figure~\ref{fig:months}(a, b, c). 

 Figures~\ref{fig:r1-r1} to~\ref{fig:r3-r3} show examples of some of the graphs for other combinations. For example, Figure~\ref{fig:r1-r1} shows the origin of sea ice retreat events (green dots) in the Southern Ocean during December 8-14, 2019. The graph further depicts how these events propagate through the sea ice grid before ultimately reaching and influencing ice shelf melting (red dots). As a reminder, ice shelve are floating platforms of glacial ice that extend from land-based ice sheets or AIS or glaciers over the cold coastal ocean. Green nodes represent sea ice retreat regions, while red nodes indicate ice shelf melting regions. The background contour outlines the AIS edge, highlighting areas where retreating sea ice is spatially linked to inland AIS melt. This visualization clearly demonstrates a connectivity where sea ice retreat events lead to, and affect, ice shelf melting.  This is shown for a linkage between Median-Q3 sea ice retreat events to Median-Q3 ice shelf retreat events. Please note that ice shelve often extend from the land to the ocean and come into contact with the ocean. Hence, they can be observed over an oceanic grid near the continental boundary. Figure~\ref{fig:r1-r3} shows the same but for linkages between medium (median - Q3) sea ice retreat and anomalous (>UB) ice shelve melting. All other plots show the combinations of linkages for different sea ice retreat and ice shelve melting scenarios. These results show how graph theory can be used to decipher the spatial linkages between two scientific areas of focus. It appears that the sea ice, experiencing different levels of retreat, gradually affects nearby oceanic grids while progressing towards the land. Please note that these changes take time to occur, as we are able to identify the spatial linkage over the weekly scale. For example, Figure~\ref{fig:r1-r1} shows the linkages (ocean to ocean and finally ocean to land) that occurred between December 8-14, 2019. We have also explored on a daily scale, but it appeared that these progressions and linkages take time to occur. A weekly scale is suitable for these linkages to be established between the start node over the ocean and the end node over the land. Thus, weekly linkages for all other combinations are shown as examples in Figures~\ref{fig:r1-r1} to~\ref{fig:r3-r3}. To avoid redundancy, we do not show graphs for all the combinations.

\subsection{CMAD Based Spatial-Link Results}

We now turn our attention to the results derived from the Convolution Matrix Anomaly Detection (CMAD)~\cite{devnath2024cmad}, which is designed to capture spatially coherent patterns of co-occurring sea ice retreat and AIS melting. To comprehensively assess the spatial dynamics of cryospheric linkages during December 2019, we have categorized AIS melt levels into three progressive threshold bands: moderate (Median–Q3), severe (Q3–UB), and extreme (>UB). The analysis reveals how regions with anomaly intensities in sea ice retreat and AIS melting with different levels exhibit varying degrees of spatial connectivity, ranging from persistent moderate interaction zones to dynamic extreme melt clusters. The following subsections detail these findings and highlight the potential linkage between CMAD-based sea ice retreat and AIS melting to expose nuanced feedback mechanisms in AIS melting in the Antarctic cryosphere. Please note that the CMAD only detects anomalous sea ice retreat events, thus, these linkages show the linkages between anomalous sea ice retreat events to three different types of ice shelf melting events. 

\subsubsection*{CMAD-Based sea ice retreat and AIS melt linkage Patterns}

The CMAD-based evaluation of moderate anomalous sea ice retreat and AIS melt (within the Median-Q3 threshold range) for December 2019 revealed persistently linked cryospheric zones across three successive weeks.  Major clusters included segments from Marie Byrd Land ($119.75^\circ$W to $110.25^\circ$W), the eastern Amundsen Sea ($107.75^\circ$W to $98.5^\circ$W), and the Bellingshausen sector ($75.0^\circ$W to $62.25^\circ$W, $69.25^\circ$S to $63.75^\circ$S). Additional linkages have been found along the Antarctic Peninsula and Getz Ice Shelf regions (e.g., $52.0^\circ$W to $46.25^\circ$W, $77.75^\circ$S to $76.75^\circ$S). These repeated spatial patterns indicate coherent environmental forcing driving moderate sea ice retreat and AIS melt co-variability across Western Antarctica~\cite{thurnherr2014export}, validating the CMAD-based Spatial-Link capability to detect nuanced spatial linkages overlooked by conventional techniques.

In the CMAD-based evaluation of anomalous sea ice retreat and AIS melt (within the Q3–UB threshold range), several spatially linked interaction zones have been identified across December 2019 and are shown in Figure~\ref{fig:months_cmad}(b). These results show a wider spread of the linkages of the two entities over similar locations.  For the extreme anomaly category where anomalous sea ice retreat and AIS melt exceed their respective upper bounds (>UB), the CMAD-based Spatial-Link has identified five concentrated and spatially connected melting zones during early December 2019, which are shown in Figure~\ref{fig:months_cmad}(c). 

It is to be noted that while we observed almost no linkages in Figures~\ref{fig:months}(g-i) when we have explored the role of sea ice retreat events with retreating intensities > UB on ice shelve melting, we have observed several linkages when we consider anomalous sea ice retreat events from CMAD and ice shelve melting in Figure~\ref{fig:months_cmad}. This is because CMAD uses the discrete convolutional operation of CNN with kernels of 2 $\times$ 2 and employs the inverse max-pooling operation over the convolved grids to identify the clusters of the anomalous retreat events. In contrary, the sea ice retreat events have been detected and used to identify the linkages in Figures~\ref{fig:months}(g-i) are based on individual grids. Thus, the process underscores the detections of sea ice retreat events in Figures~\ref{fig:months}(g-i). It shows the importance and suitability of using machine learning methods (such as the convolution operation in CMAD)  to identify anomalies rather than traditional statistical methods. 
Figures~\ref{fig:r2} and~\ref{fig:r3} illustrate representative graphs for each threshold combination (Q3–UB and >UB), highlighting spatially coherent sea ice–ice shelf linkage zones.
 
\subsection{Spatial-Link$_{Benchmark}$ Results}

For benchmarking, we have applied the Spatial-Link$_{Benchmark}$ to sulfate aerosol transport events during September 2020 to evaluate its ability to identify statistically significant spatial linkages or paths corresponding to wildfire-driven atmospheric rivers. A total of 16 time-step graphs have been constructed from MERRA-2 sulfate aerosol data for the dates 20200909, 20200912, 20200913, and 20200915, with four graphs per day corresponding to 6-hour intervals. 20200915 of the three time stamps exhibited statistically significant linkage paths between wildfire origin regions and PM$_{2.5}$ levels at stations, which is shown in Figure~\ref{fig:benchmark}. After constructing the sulfate AAR based graphs, we have analyzed whether significant spatial linkages or paths existed from wildfire regions to downwind locations experiencing elevated particulate matter. We have integrated PM$_{2.5}$ measurements from IMPROVE network's of Gate of the Mountains station, Montana~\cite{chakraborty2022aerosol}, focusing on recorded spikes that temporally aligned with MERRA-2 aerosol events. These PM$_{2.5}$ spikes have been used as target nodes for path-finding analysis, as shown in Figure~\ref{fig:benchmark}. Our analysis shows that the sulfate aerosol river originated from Southern California, following the outbreak of four major wildfires—Creek Fire, El Dorado Fire, Bobcat Fire, and Slater-Devil Fire—which ignited between September 4 and 7, 2020. The river reached the IMPROVE network measurement station that shows an increase in the PM$_{2.5}$ concentrations beyond the recommended level by the World Health Organization (WHO) 15~\textmu g\,m$^{-3}$. The concentration recorded at that station on that day was 75~\textmu g\,m$^{-3}$. This has been previously explored by Chakraborty et al. in 2023~\cite{chakraborty2023aerosol}. By using a different metric, which developed the AAR database that is publicly available for download for research purposes~\cite{AAR_22}, they show the presence of the sulfate AAR during the period of September 9th to 15th that raised the PM$_{2.5}$ concentration of the station beyond the WHO recommended human exposure level of 15~\textmu g\,m$^{-3}$. To evaluate the statistical significance of the observed spatial connections, we have performed randomized path analysis across different time snapshots. The results have revealed that the sulfate aerosol transport patterns observed in the AAR data derived from MERRA-2 reanalysis have coincided with fire events and established a graph between the starting nodes (fires) and the Gate of the Mountains station in Montana. These findings support the use of our proposed method to identify significant sulfate transport influenced by wildfires. These path has low $p$-values ($< 0.005$), indicating that the observed transport pathways have been highly unlikely to occur by chance in randomized null models. The identified paths have been aligned well with elevated PM$_{2.5}$ levels at Gates of the Mountains (GAMO1)~\cite{chakraborty2022aerosol}, confirming the physical relevance of the detected linkages. In other time steps, no valid paths have been found, further reinforcing that the method selectively identifies meaningful spatio-temporal events rather than random connections.



\section{Conclusion and Future Work }
\label{sec:con}

In this study, we have introduced a Spatial-Link framework to detect and analyze spatially coherent linkages between sea ice retreat and AIS melting across Antarctica using December 2019 as a case study. By segmenting into threshold-based categories—ranging from moderate (Median–Q3) to extreme (>UB)—our approach has successfully identified persistent and physically relevant co-occurrence patterns. These spatial linkages, observed across both coastal and inland regions, suggest that sea ice loss often precedes or coincides with intensified AIS response, especially in dynamically sensitive sectors such as the Amundsen and Bellingshausen Seas. This study bridges the gap between the sea ice and ice sheet research communities and points out that there is a need to consider the linkages to better understand the cryosphere processes to reduce uncertainty in the existing models. 

In our future work, we will expand this analysis temporally and climatologically by incorporating multi-year and seasonal datasets. In particular, we aim to (1) quantify the lead–lag relationships between sea ice retreat and AIS melting, (2) incorporate atmospheric and oceanic reanalysis variables (e.g., surface winds, sea surface temperatures, and geopotential heights) to infer causal drivers, and (3) develop a predictive model using machine learning and graph neural networks. Additionally, we plan to apply this framework to other polar regions and refine its robustness using multi-resolution satellite data. These extensions will help enhance early warning systems, support adaptive cryosphere modeling, and inform global sea-level risk assessments. Please note that an in-depth analysis is needed to find out the exact number of nodes (11 for this study), and it needs extensive evaluations using multiple years of satellite data. This study established the foundation to advance our understanding of heterogeneous cryospheric linkages and to serve as a tool for long-term climate diagnostics, policy planning, and the development of resilient adaptation strategies in the face of accelerating polar change.


\begin{acks}
This work is supported by NSF Award \#2118285, "iHARP: NSF HDR Institute for Harnessing Data and Model Revolution in the Polar Region." 
\end{acks}


\bibliographystyle{ACM-Reference-Format}
\bibliography{sample_base}

\end{document}